# Le Chatelier's Principle and Field-Induced Change in Magnetic Entropy Leading to Spin Lattice Partitioning and Magnetization Plateau

Myung-Hwan Whangbo[1,*], Hyun-Joo Koo[2], and Olga S. Volkova[3,*]

[1] Department of Chemistry, North Carolina State University, Raleigh, NC 27695-8204, USA

[2] Department of Chemistry, Research Institute for Basic Sciences, Kyung Hee University, Seoul 02447, Republic of Korea

[3] Department of Low Temperature Physics and Superconductivity, Lomonosov Moscow State University, Moscow 119991, Russia

mike_whangbo@ncsu.edu

os.volkova@yahoo.com

## Abstract

For a certain antiferromagnet, the magnetization $M(H)$ does not increase gradually with increasing magnetic field $\mu_0H$ but exhibits field region(s) where $M(H)$ remains constant typically at an integer fraction of its saturation magnetization $M_{sat}$. This phenomenon is readily understood by the supposition that, under external magnetic field, such an antiferromagnet undergoes field-induced partitioning of its spin lattice into ferrimagnetic fragments. We searched for a theoretical basis for this supposition by investigating how external magnetic fields affect the magnetic entropy of such an antiferromagnet, to find that the field region of the magnetization plateau has a single magnetic phase, but a nonzero slope region of the magnetization curve has two magnetic phases of different magnetic entropy. Our analysis predicts that the magnetic entropy of a single-phase region does not depend on magnetic field but that of a two-phase region does such that the magnetic entropy increases with field if the two phases dynamically shift their spin sites but decreases if their spin sites remain fixed. We tested these predictions by carrying out magnetization and specific heat measurements for $\gamma$-Mn$_3$(PO$_4$)$_2$, which exhibits a 1/3-magnetization plateau at 2 K at magnetic fields between ~6 and ~23 T and undergoes a long-range antiferromagnetic ordering at $T_N$ = 12.9 K. It was found that the magnetic entropy of the two-phase region increases with field, indicating that field-induced breaking of magnetic bonds and hence field-induced partitioning of an antiferromagnetic spin lattice are time-averaged results of all allowed spin arrangements that occur repeatedly during static magnetization measurements. The temperature-dependent magnetic specific heats $C_m(T)$ of $\gamma$-Mn$_3$(PO$_4$)$_2$ between 2 – 6 K shows a larger excitation gap $\Delta$ when measured at 9 T than at 0 T (i.e., $\Delta$ = 1.4 vs. 0.5 K). These energy gaps reflect the two successive local excitations of linear Mn$^{2+}$-Mn$^{2+}$-Mn$^{2+}$ ferrimagnetic trimers embedded in the antiferromagnetic spin lattice of $\gamma$-Mn$_3$(PO$_4$)$_2$ and arise from the Boltzmann factor associated with these excitations. Our work demonstrates that Le Chatelier's principle provides a qualitative basis for understanding a series of events that an external magnetic field can bring about in antiferromagnets exhibiting magnetization plateaus.

## 1. Introduction

Magnetization plateaus, i.e., regions of constant magnetization despite increasing external magnetic field, observed for various antiferromagnets are readily understood by the supposition[1] that an antiferromagnet counteracts the field according to Le Chatelier's principle by absorbing Zeeman energy. This requires partitioning of its spin lattice into ferrimagnetic fragments. For a magnetic species of spin $S$, the spin moment $\mu_s$ is given by $\mu_s = -\mu_B g S$. Under magnetic field $\mu_0 H$, such a magnetic species has the Zeeman energy $E_Z$ given by

$$E_Z = -\mu_0 \vec{\mu}_s \cdot \vec{H} = \mu_0 \mu_B g \vec{S} \cdot \vec{H} \qquad (1)$$

Thus, the more ferrimagnetic fragments an antiferromagnet generates, the more Zeeman energy it absorbs. The Zeeman energy of an individual ferrimagnetic fragment increases with magnetic field, and the accumulated Zeeman energy creates additional ferrimagnetic fragments until the whole spin lattice is partitioned into ferrimagnetic fragments, hence reaching the magnetization plateau. An important implication of this picture is that, while the spin lattice undergoes partitioning out ferrimagnetic fragments, it consists of two different magnetic phases, namely, one made up of partitioned-out ferrimagnetic fragments and the other free of such fragments.

To illustrate the concept of partitioning an antiferromagnetic (AFM) spin lattice, we first consider a simple antiferromagnet with no spin frustration, e.g., a chain of antiferromagnetically coupled ferrimagnetic trimers (**Fig. 1a**) in which the intra-trimer exchange J and the inter-trimer exchange J′ are both AFM with J stronger than J′. (Here we use the convention that AFM spin exchanges are represented by positive spin exchanges.) In the ground state spin arrangement of this chain, the trimers are antiferromagnetically coupled as depicted in **Fig. 1a**. Under magnetic field, this chain starts to break the weak magnetic bonds J′ to generate ferrimagnetic trimers, enabling the spin lattice to absorb Zeeman energy until all inter-trimer bonds are broken (**Fig. 1b**), where each broken inter-trimer bond is represented as ferromagnetically coupled (see below for further discussion). The situation becomes different for a spin-frustrated spin lattice like a trigonal one (**Fig. 1c**) described by the nearest-neighbor spin exchange J. The 1/3-magnetization plateau observed for such a spin lattice is readily explained by supposing that the field partitions the spin lattice into ferrimagnetic triangles (**Fig. 1d**).[1] However, this spin lattice has no weak magnetic bonds to break but is most likely in a fluctuating liquid-like ground state. On increasing magnetic field, this spin lattice (**Fig. 1c**) would become increasingly spin polarized hence weakening the fluctuation toward the structure composed of partitioned-out ferrimagnetic triangles (**Fig. 1d**). In principle, one may consider four different mechanisms concerning how the process of field-induced spin polarization takes place when the field is increased by assuming that a trigonal spin lattice consists of *n* non-overlapping triangles.

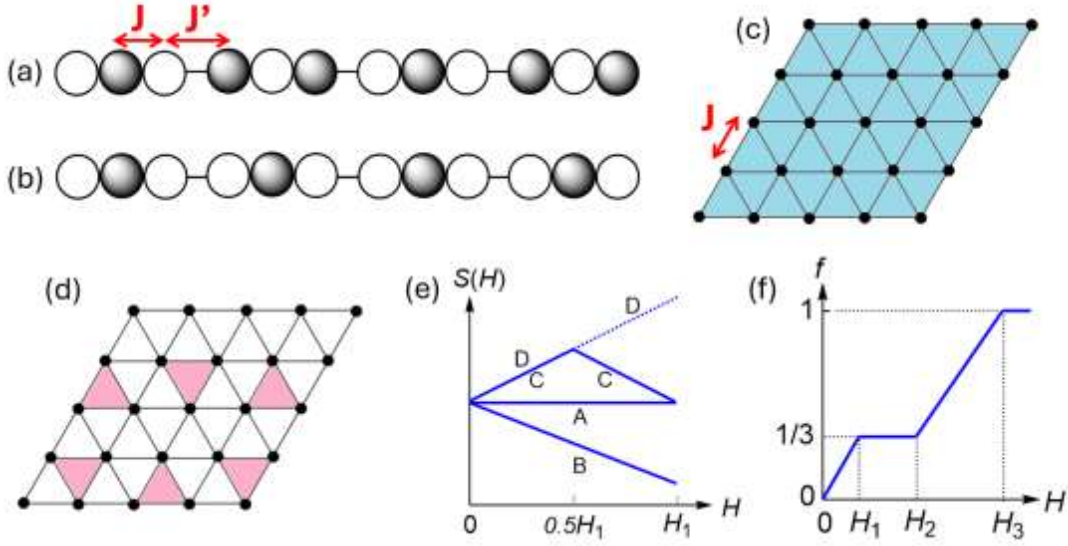

**Fig. 1**. (a) AFM chain of antiferromagnetically coupled ferrimagnetic trimers. (b) Ferrimagnetic chain of ferromagnetically coupled ferrimagnetic trimers, which results from the AFM chain by breaking all inter-trimer bonds. (c) Trigonal spin lattice. (d) Trigonal spin lattice partitioned into nonoverlapping ferrimagnetic triangles under magnetic field. (e) Change in the magnetic entropy $S(H)$ during the process of field-induced spin polarization arising from Models A – D showing schematically whether it increases, decreases or remains constant with increasing field. For simplicity, a linear change was assumed. (f) Schematic magnetization curve expected for a trigonal spin lattice, where f represents a fraction of the saturation magnetization $M_{sat}$ defined as f = $M/M_{sat}$.

Model A:

The spin polarization process is homogeneous throughout the spin lattice. The magnetic structure of the spin lattice changes to increase the degree of spin polarization with increasing field. However, at a given magnetic field, there is only one magnetic configuration which is uniform throughout the spin lattice. Thus, the magnetic entropy of the spin lattice remains zero as the field increases.

Model B:

The spin polarization process is statically heterogeneous in the spin lattice so that, once generated, the partitioned-out ferrimagnetic fragments have their spin sites remain fixed in the spin lattice, and the degree of spin polarization increases with increasing the number of partitioned-out ferrimagnetic fragments. Provided that a given field generates $m$ ferrimagnetic fragments, the total number of different ways to choose the positions of these fragments in the given spin lattice is given by the binomial coefficient, $_n C_m \equiv \Omega(m)$. However, only one of $\Omega(m)$ is chosen at a given field. As $m$ increases to $m+1$ with increasing field, one of $\Omega(m+1)$ will be chosen which is consistent with the chosen one of $\Omega(m)$. This is akin to a crystallite growing from a single nucleation site in the melt. The spin degree of freedom is lower in the partitioned-out ferrimagnetic fragments than in the spin-unpolarized fragments in the rest of the spin lattice, so the overall magnetic entropy of the spin lattice decreases with increasing field.

Model C:

The spin polarization process is dynamically heterogeneous in the spin lattice so that the partitioned-out ferrimagnetic fragments are separated from the rest of the spin lattice free of such fragments, and the degree of spin polarization increases with increasing the number of partitioned-out ferrimagnetic fragments. These aspects are identical with those of Model B. However, Model C allows the spin sites of the partitioned-out fragments in the spin lattice to shift dynamically due to spin fluctuations. Thus, the $m$ ferrimagnetic fragments generated at a given field can adopt all possible ways allowed by $\Omega(m)$ by dynamically shifting their spin sites rapidly. Then, the configurational magnetic entropy $S(m)$ is given by $S(m) = k_B ln\Omega(m)$. Let $\mu_0 H_1$ be the onset field of the magnetization plateau where the spin lattice is solely composed of the ferrimagnetic fragments. Then, as the field increases from 0 to $\mu_0 H_1$, $m$ increases from 0 to $n$. The binomial coefficient $\Omega(m)$ is a symmetric function of $m$; $\Omega(m)$ increases from 1 to the maximum, $\Omega(n/2)$, as $m$ increases from 0 to $n/2$, but it decreases from this maximum to 1 as $m$ increases from $n/2$ to $n$.

Model D:

This model is the same as Model C except that the total number of different ways to choose the positions of $m$ partitioned-out fragments is given by modified binomial coefficients, which we write as $\langle {}_n C_m \rangle \equiv \langle \Omega(m) \rangle$ (see below). The latter increases gradually as $m$ increases from 0 to $n$ so that the associated magnetic entropy, $\langle S(m) \rangle = k_B ln \langle \Omega(m) \rangle$, increases steadily as $m$ increases from 0 to $n$ (see below).

In summary, we note that, as the field increases from 0 to $\mu_0 H_1$, the magnetic entropy $S(H)$ remains constant in Model A, but decreases gradually in Model B. In Model C, $S(H)$ increases with field from 0 to $\mu_0 H_1/2$ but decreases with from $\mu_0 H_1/2$ to $\mu_0 H_1$, following the behavior of the binomial coefficient. In Model D, $S(H)$ increases with field from 0 to $\mu_0 H_1$.

The magnetization $M(H)$ of the trigonal spin lattice as a function of the magnetic field is schematically depicted in **Fig. 1f**. It increases almost linearly with field $\mu_0 H$ as the field increases from 0 to $\mu_0 H_1$ (i.e., the onset field of the 1/3-magnetization plateau). In terms of Models B – D, this is achieved by increasing the number of partitioned-out ferrimagnetic triangles while the remaining areas of the spin lattice are unaffected by the field. Thus, the nonzero slope region is a two-phase region, a static one in Model B but a dynamically fluctuating one in Model C and D. As the field increases from $\mu_0 H_1$ to $\mu_0 H_2$, the magnetization does not change giving rise to a 1/3-magnetization plateau because the field is not strong enough to convert each ferrimagnetic triangle to a fully polarized (i.e., ferromagnetic) one. (See below for further discussion on the magnetization curve beyond $\mu_0 H_2$.) In the field region of the magnetization curve represented by a nonzero slope (e.g., between 0 and $\mu_0 H_1$ as well as between $\mu_0 H_2$ and $\mu_0 H_3$), two different magnetic phases coexist within the same spin lattice. The spin lattice of any magnet is accommodated by its crystal lattice which, apart from weak magnetoelastic effects, remains structurally intact throughout the magnetization measurements so that the spin exchanges of the magnet remain essentially unchanged throughout the magnetization process. Thus, it is important to understand how the two magnetic phases differ but coexist in the same spin lattice, and hence what is meant by the field-induced partitioning of a spin lattice into ferrimagnetic fragments, which involves the breaking of magnetic bonds.

In the present work we explore the question raised above, to find that the magnetic entropy of an AFM spin lattice leading to a magnetization plateau does not depend on field in the region of a nonzero magnetization plateau, but it does in other regions of the magnetization curve. The predictions of our theoretical analysis were tested by measuring the temperature- and field-dependent specific heat of $\gamma$-Mn$_3$(PO$_4$)$_2$,[2] which exhibits a 1/3-magnetization plateau between ~6 and ~23 T. Our work is organized as follows: In Section 2, we present the results of our theoretical analysis on how external magnetic field affects the magnetic entropy of an antiferromagnet. Section 3 describes the temperature- and field-dependence of the specific heat measured for $\gamma$-Mn$_3$(PO$_4$)$_2$ as an example for testing the theoretical predictions of Section 2. After discussing several important implications of our work in Section 4, we summarize our conclusions in Section 5.

## 2. Results: Theoretical Analysis

In this section our discussion is based on a trigonal antiferromagnet with trigonal spin lattice, to find that the field-dependence of magnetic entropy and the nature of the magnetic phase in the field region describing the magnetization plateau differ from those describing the nonzero slope regions of the magnetization curve. These conclusions remain valid for other antiferromagnets exhibiting the magnetization plateau phenomenon even if their spin lattices are not trigonal (see below).

### 2.1. Field-induced change in magnetic entropy

As discussed above, the spin lattice of an antiferromagnet exhibiting a magnetization plateau is partitioned into two different magnetic phases under field. To understand how the two magnetic phases differ and clarify what is meant by the field-induced breaking of magnetic bonds, we consider a trigonal spin lattice consisting of N spin sites (i.e., $n = $ N/3 nonoverlapping triangles) with AFM nearest-neighbor spin exchange J. In the two-phase region of the magnetization curve, we assume that the N spin sites are divided into the M and N − M sites. The M region contains $m$ = M/3 partitioned-out, ferrimagnetic, and nonoverlapping triangles. Some examples of $m = 1 - 4$ are presented in **Fig. 2a − 2d**, respectively, where the partitioned-out ferrimagnetic triangles are represented by pink triangles, the spin-unpolarized triangles by cyan triangles, and the triangles forming the boundary between the M and N – M regions by unshaded triangles. Note that the partitioned-out ferrimagnetic triangles can be anywhere in the spin lattice, and they may occur in separate places or can be adjacent to each other.

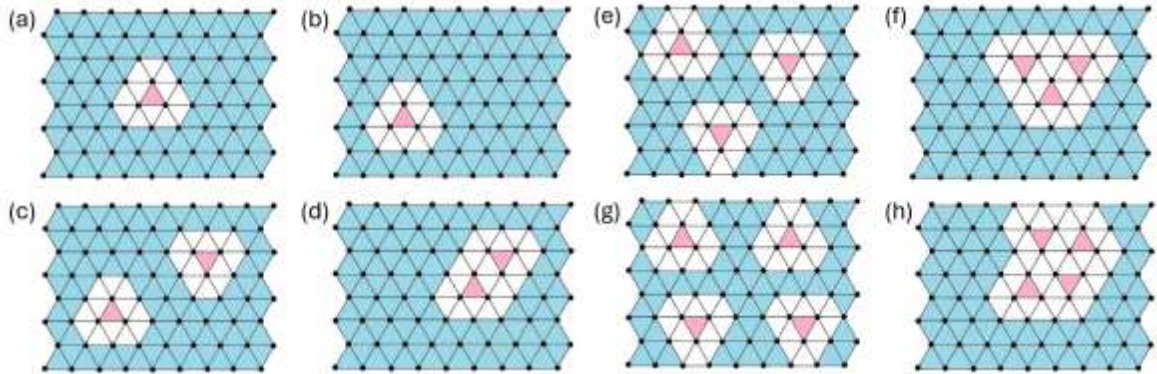

Fig. 2. (a, b) Two examples of one partitioned-out ferrimagnetic triangle. (c, d) Two examples of two partitioned-out ferrimagnetic triangles. (e, f) Two examples of three partitioned-out ferrimagnetic triangles. (g, h) Two examples of four partitioned-out ferrimagnetic triangles.

To examine the magnetic entropy associated with the process of field-induced spin polarization, we first consider each triangle by labeling its three vertices as 1, 2 and 3 (**Fig. 3a**), and represent the up-spin ↑ (down-spin ↓) at each vertex by the letter u (d). Then, the ↑↓↑ (udu) spin arrangement is depicted as in **Fig. 3b**, and for the ↑↑↓ (uud) spin arrangement as in **Fig. 3c**.

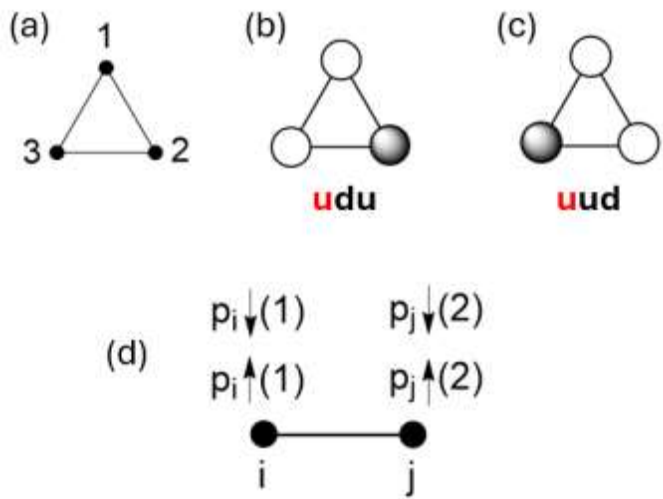

**Fig. 3**. (a) Triangle with spin sites 1, 2 and 3. (b) The ↑↓↑ (udu) spin arrangement. (c) The ↑↑↓ (uud) spin arrangement. The up-spin (↑) and down-spin (↓) are represented by unshaded and shaded circles, respectively, and by letters u and d, respectively. The up spin at site 1 is represented by red label u to indicate that it remains the same in the two arrangements. (d) Nearest-neighbor spin exchange path i-j between spin sites i and j for the case when they belong to the triangles of magnetic phases 1 and 2, respectively. Here $p_i\uparrow(1)$ [$p_i\downarrow(1)$] refers to the number of up-spin (down-spin) possibilities at the site i of phase 1. Similarly, $p_i\uparrow(2)$ [$p_i\downarrow(2)$] refers to the number of up-spin (down-spin) possibilities at the site j of phase 2.

Then, each ferrimagnetic triangle of the M region can have the following six arrangements:

uud, udu; uud, duu; udu, duu,

each of which has net one up-spin. Let $p_i\uparrow$ ($p_i\downarrow$) be the number of up-spin (down-spin) possibilities at each site i (= 1, 2 or 3) of each partitioned-out ferrimagnetic triangle. Then, as discussed above, $p_i\uparrow = 3$ and $p_i\downarrow = 2$ in the M region. In the N − M region, each triangle can have the 12 possible spin arrangements:

uud, udu; uud, duu; udu, duu

dud, ddu; udd, ddu; udd, dud

where each of the first six arrangements has net one up-spin, while each of the second six arrangements has net one down-spin. Thus, in the (N − M) region, each site i of a given triangle has $p_i\uparrow = p_i\downarrow = 5$ and hence does not contribute to the magnetization. Consequently, the field-induced partitioning of a trigonal spin lattice into partitioned-out ferrimagnetic triangles amounts to the conversion from the spin-unpolarized triangles of the 12 possible spin arrangements to the ferrimagnetic triangles of the six possible spin arrangements. This means that the magnetic entropy per spin site is lower in the M region than in the N − M region by a factor of 2. Thus, if Model B is the correct process of field-induced spin-polarization, the magnetic entropy would decrease gradually with increasing field. However, if Model C or D is correct, the opposite is predicted because each partitioned-out ferrimagnetic triangle can be anywhere in the spin lattice. As already discussed, the configurational magnetic entropy is given by $k_B ln\Omega(m)$ in Model C, and by $k_B ln\langle\Omega(m)\rangle$ in Model D (**Fig. 1e**). Due to this configurational entropy, the overall magnetic entropy would be much greater in the nonzero slope region (0 to $\mu_0 H_1$ in **Fig. 1f**) than in the 1/3-magnetization plateau region ($\mu_0 H_1$ to $\mu_0 H_2$ in **Fig. 1e**), because the latter region consists of only the partitioned-out ferrimagnetic triangles described by only one spin configuration. With further increasing the field (from $\mu_0 H_2$ to $\mu_0 H_3$ in **Fig. 1f**), ferrimagnetic triangles start to become fully magnetized (i.e., fully spin-polarized) with only one spin arrangement, uuu, for each triangle. The latter creates another two-phase region in the magnetization curve, where the fully magnetized triangles coexist with ferrimagnetic triangles. If Model C or D is correct, the overall magnetic entropy is greater in this nonzero slope region ($\mu_0 H_2$ to $\mu_0 H_3$ in **Fig. 1f**) than in the completely magnetized region above $\mu_0 H_3$ where $p_i\uparrow = 1$ and $p_i\downarrow = 0$.

In short, the magnetic entropy of a spin lattice remains constant in the field region of a single phase. In the field region of a two phase, the magnetic entropy of a spin lattice decreases with field if Model B is correct, but this is not the case if Model C or D is correct.

## 2.2. Field-induced breaking of magnetic bonds

In this section we examine the field-induced breaking of magnetic bonds from the viewpoint of magnetic entropy. The two-phase region between 0 and $\mu_0 H_1$ (**Fig. 1f**) of the magnetization curve consists of spin-unpolarized triangles and ferrimagnetic triangles (**Fig. 2**). In this region, broken magnetic bonds are the spin exchange paths bridging these two different phases. The two-phase region between $\mu_0 H_2$ and $\mu_0 H_3$ (**Fig. 1f**) of the magnetization curve consists of ferrimagnetic triangles and fully spin-polarized triangles. In this region, broken magnetic bonds are the spin exchange paths bridging these two different phases.

Let us first consider the field region between 0 and $\mu_0 H_1$. For each site i of a ferrimagnetic triangle (say, phase 1), $p_i\uparrow(1) = 3$ and $p_i\downarrow(1) = 2$. For each site j of spin-unpolarized triangle (say, phase 2), $p_j\uparrow(2) = p_j\downarrow(2) = 5$. Then, the allowed spin arrangements for the spin exchange path i–j bridging between adjacent ferrimagnetic and spin-unpolarized triangles (i.e., each broken magnetic bond between) can be divided into the $p_i\uparrow(1)-p_j\uparrow(2)$ and $p_i\uparrow(1)-p_j\downarrow(2)$ as well as the $p_i\downarrow(1)-p_j\uparrow(2)$ and $p_i\downarrow(1)-p_j\downarrow(2)$ arrangements (see **Fig. 3d**). These lead to 25 ways of AFM coupling and 25 ways of ferromagnetic (FM) coupling between the sites i and j. Effectively, then, there is net no spin exchange interaction in the magnetic bond i–j, as if there is no structural connection between the spin sites i and j. This provides a theoretical basis for the supposition of field-induced breaking of magnetic bonds and hence the supposition of field-induced partitioning of an AFM spin lattice into ferrimagnetic fragments.

Consider now the interaction between two adjacent ferrimagnetic triangles in the field region between 0 and $\mu_0H_1$ (**Fig. 1f**). The spin arrangements between the spin site i in one ferrimagnetic triangle (say, phase 1) and the spin site j of its adjacent ferrimagnetic triangle (phase 1) are divided into the $p_i\uparrow(1)-p_j\uparrow(1)$ and $p_i\uparrow(1)-p_j\downarrow(1)$ as well as the $p_i\downarrow(1)-p_j\uparrow(1)$ and $p_i\downarrow(1)-p_j\downarrow(1)$ arrangements. These four arrangements lead to 13 ways of FM coupling and 12 ways of AFM coupling. In other words, net one out of 25 ways results in FM coupling, which gives rise to the destabilization of $(1/25)J$ per such i-j contact. In the field region between $\mu_0H_2$ and $\mu_0H_3$ (**Fig. 1f**), each partitioned-out FM triangle (say, phase 1) has $p_i\uparrow(1) = 1$ and $p_i\downarrow(1) = 0$. For each ferrimagnetic triangle (say, phase 2), $p_j\uparrow(2) = 3$ and $p_j\downarrow(2) = 2$, as already noted. Then, the allowed spin arrangements for each spin exchange path i–j between adjacent FM and ferrimagnetic triangles are divided into the $p_i\uparrow(1)-p_j\uparrow(2)$ and $p_i\uparrow(1)-p_j\downarrow(2)$ arrangements, which leads to three ways of FM coupling and two ways of AFM coupling. Thus, one out of five ways results in FM coupling, so each i-j contact causes the destabilization of $(1/5)J$. For each magnetic bond i-j between two adjacent FM triangles, there is only one way of FM coupling, leading to the destabilization of J per such i-j contact.

In the field region between 0 and $\mu_0H_1$, therefore, it is energetically more favorable to surround a ferrimagnetic triangle with spin-unpolarized triangles rather than with ferrimagnetic triangles by $(1/25)J$ per i-j contact. Similarly, in the field region between $\mu_0H_2$ and $\mu_0H_3$, it is energetically more favorable to surround an FM triangle with ferrimagnetic triangles than with FM triangles by $(4/5)J$ per i-j contact. These observations have important implications as will be discussed later.

Our discussions described above are readily extended to antiferromagnets with no spin frustration, providing theoretical support for the supposition of field-induced breaking of magnetic bonds and partitioning of their AFM spin lattice into ferrimagnetic fragments. These were presented in Section S1 with **Fig. S1** and **Fig. S2** in the Supporting Information by considering an AFM chain of antiferromagnetically coupled ferrimagnetic trimers.

The above discussion implicitly assumed that all spin arrangements between any two adjacent spin sites are equally probable and occur during static magnetization measurements. This assumption is very reasonable, given that the time scale of spin fluctuations ($10^{-12}$ to $10^{-4}$ seconds) is fast. Then, the field-induced breaking and hence the field-induced partitioning of an antiferromagnetic spin lattice should be understood as the time-averaged results of all allowed spin arrangements that can occur repeatedly during the static magnetization measurements. The field-induced breaking of magnetic bonds in an AFM spin lattice results from field-induced dynamically heterogeneous spin-polarizations, and the broken magnetic bonds are the magnetic bonds bridging the lower- and higher-entropy phases, which effectively have no spin exchange interaction as if physically broken. These arguments are supported if the process of field-induced spin-polarization is governed by Model C or D. For experimental support for this conclusion, we examine the field-dependence of the specific heat of $\gamma$-Mn$_3$(PO$_4$)$_2$ in the next section.

### 3. Results: Specific heat measurements

To test the conclusion that the magnetic entropy of an antiferromagnet exhibiting a nonzero-magnetization plateau does not depend on field in the region of the nonzero magnetization plateau, but it does in other regions of the magnetization curve we explored the temperature- and field-dependence of the specific heat measured for an antiferromagnet with a nonzero-magnetization plateau. $\gamma$-Mn$_3$(PO$_4$)$_2$, consisting of Mn$^{2+}$ (d$^5$, $S$ = 5/2) ions, is a suitable system to investigate because its 1/3 magnetization plateau (between ~6 and ~23 T at 2 K)[2] lies in the easily accessible regime of standard laboratory equipment and its three-dimensional (3D) AFM ordering temperature $T_N$ = 12.9 K[2] lies well above the typical temperature of magnetization measurements.

### 3.1. Spin lattice of $\gamma$-Mn$_3$(PO$_4$)$_2$

For the convenience of our discussions in the following, we briefly review the essential features of the spin lattice of $\gamma$-Mn$_3$(PO$_4$)$_2$. The spin lattice consists of layers made up of linear trimers Mn1-Mn2-Mn1 in which the central Mn2 makes two Mn2-Mn1 bridges with the end Mn1 atoms of two different trimers (**Fig. 4a**). Such layers are interconnected by the Mn1-Mn1 bridges as shown in **Fig. 4b**, which depict a side-projection view of three adjacent layers indicated by rectangular boxes.

The intra-trimer Mn2-Mn1 spin exchange ($J_3$), the inter-trimer Mn2-Mn1 spin exchange ($J_2$) within each layer, and the inter-trimer spin exchange ($J_1$) between adjacent layers are all AFM ($J_1 = 1.7$ K, $J_2 = 4.7$ K and $J_3 = 10.5$ K).[1,2] Thus, each linear trimer makes a ferrimagnetic unit (**Fig. 4c**), and every trimer of one layer is antiferromagnetically coupled with one trimer in each adjacent layer by the spin exchange $J_1$ (**Fig. 4c**). Each partitioned-out ferrimagnetic layer, represented in **Fig. 4d** by an isolated trimer, loses these interlayer magnetic bonds. In every ferrimagnetic layer of linear ferrimagnetic trimers, each trimer is antiferromagnetically coupled to four different trimers through the $J_2$ paths (**Fig. 4e**), so that every layer of linear ferrimagnetic trimers is ferrimagnetic. Such ferrimagnetic layers couple antiferromagnetically via interlayer spin exchanges $J_1$ to form a 3D AFM structure at $T_N = 12.9$ K. The 1/3-magnetization plateau of γ-$Mn_3(PO_4)_2$ results when all interlayer magnetic bonds $J_1$ are broken, which occurs in the magnetization process when the field reaches ~6 T. This plateau ends when the field increases beyond ~23 T, where the ferrimagnetic trimers start to become fully magnetized.

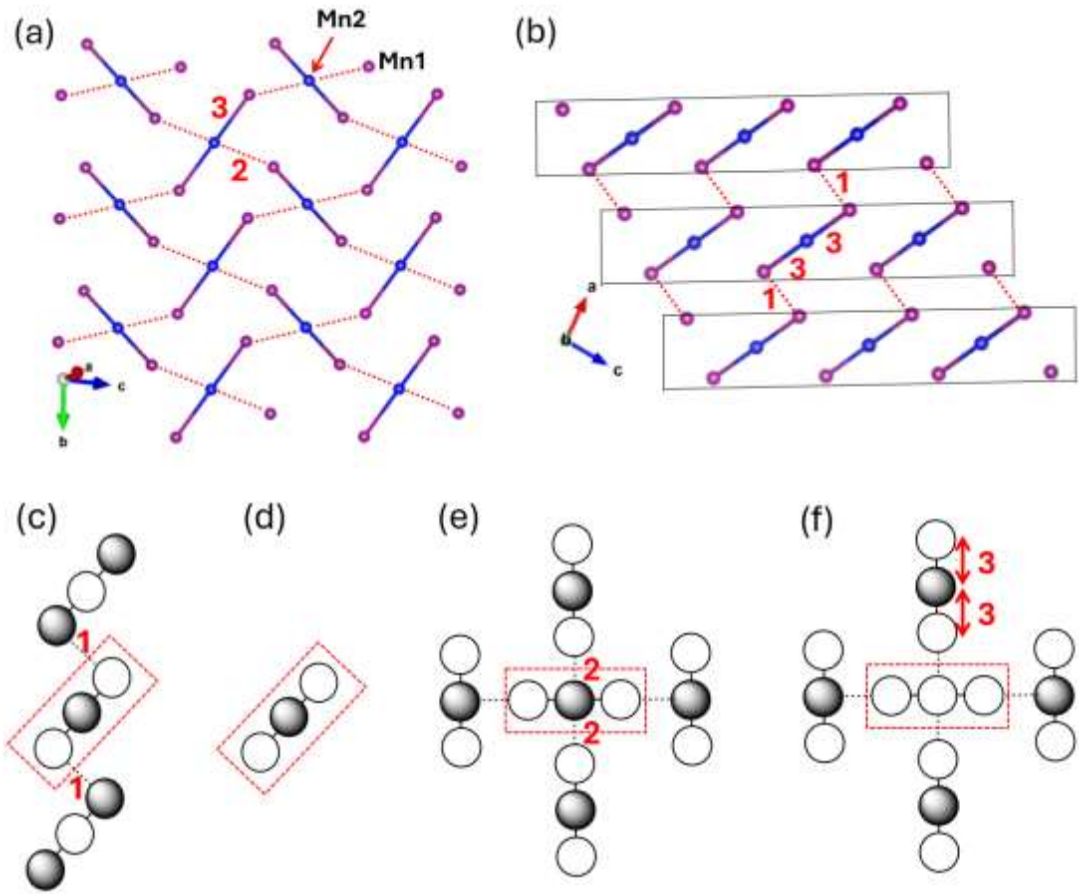

**Fig. 4**. (a). A layer of linear Mn1-Mn2-Mn1 trimers with every Mn2 making two Mn2-Mn1 bridges with its two adjacent trimers. (b) Adjacent layers of linear Mn1-Mn2-Mn1 trimers connected by Mn1-Mn1 bridges. (c) Each linear trimer of one ferrimagnetic layer antiferromagnetically coupled to one trimer in each adjacent layer via the spin exchange $J_1$. For simplicity, each ferrimagnetic layer is represented by a trimer. (d) One partitioned-out ferrimagnetic layer represented by a trimer. (e) In every ferrimagnetic layer each trimer is antiferromagnetically coupled to four adjacent ferrimagnetic trimers through the $J_2$ paths. Note that this ferrimagnetic layer is represented by a trimer in (c, d). (f) Locally excited state of a ferrimagnetic trimer in a ferrimagnetic layer made up of ferrimagnetic trimers. The red labels 1 – 3 in (a – f) refer to the spin exchanges $J_1$ – $J_3$, respectively. The unshaded (shaded) circles in (c – f) represent up-spin (down-spin) $Mn^{2+}$ sites in each linear trimer.

In summary, the spin lattice of γ-$Mn_3(PO_4)_2$ is 3D in character, in which ferrimagnetic layers are antiferromagnetically coupled with no spin frustration. The trigonal spin lattice discussed in the previous section is two-dimensional (2D) in character, in which the nearest-neighbor spin exchange is AFM so that there occurs spin frustration. In the nonzero slope region of the curve prior to the 1/3-magnetization plateau, the role of the partitioned-out ferrimagnetic layers play in γ-$Mn_3(PO_4)_2$ is analogous to that of the partitioned-out ferrimagnetic triangles do in a trigonal layer.

### 3.2. Specific heat of γ-Mn₃(PO₄)₂

We undertook specific heat measurements for γ-$Mn_3(PO_4)_2$ as a function of temperature (between 2 and 6 K) at 0 and 9 T, and as a function of magnetic field (between 0 and 9 T) at 2 K. Our measurements were carried out using a pressed pellet of γ-$Mn_3(PO_4)_2$ weighting 2.7 mg by "Quantum Design" Physical Properties Measurements System PPMS-9T with steps of 0.1 T at a constant temperature of 2 K. The specific heat was calibrated independently to find that its variation with the magnetic field was negligible.

### 3.2.A. Temperature-dependence of specific heat

In general, the temperature-dependent specific heat $C(T)$ of a magnet has two contributions, $C_{ph}(T)$ and $C_m(T)$, which are the phonon (crystal lattice vibrations) and magnon (magnetic subsystem excitations) contributions, respectively, so that $C(T) = C_{ph}(T) + C_m(T)$. Usually, at low temperatures the magnon contribution strongly prevails over the phonon contribution. At low temperatures, the $C_m(T)$ of a 3D antiferromagnet varies as ~$T^3$, and that of a 3D ferromagnet or a ferrimagnet as ~$T^{3/2}$.[3] As found for $A_2Ni_2TeO_6$ (A = K, Li),[4] an external magnetic field slightly lowers the $T_N$ and reduces slightly the $\lambda$-type anomaly because it has the effect of suppressing an AFM order. The temperature-dependent specific heat $C(T)$ of $\gamma$-$Mn_3(PO_4)_2$ (**Fig. 5**) differs noticeably from those of $A_2Ni_2TeO_6$ (A = K, Li) (**Fig. S3** in Section S2). Under zero field $\gamma$-$Mn_3(PO_4)_2$ shows a pronounced $\lambda$-type anomaly at $T_N$ (= 12.9 K), but external magnetic field broadens the $\lambda$-type anomaly and raises the $T_N$ (**Fig. 5**). This is most likely because the external field enhances the ferrimagnetic character of each layer by increasing its overall moment, which is accompanied by a moment increase at the Mn1 sites. The latter would strengthen the interlayer AFM spin exchange $J_1$, which in turn raises $T_N$. The field-induced broadening of the $\lambda$-type anomaly can occur when the field-induced enhancement of the moment is nonuniform throughout the ferrimagnetic layers. We estimate the magnetic entropy $C_m(T)$ at $\mu_0H = 0$ T by approximating the phonon contribution $C_{ph}(T)$ as the sum of two Einstein modes with $\Theta_{E1} = 210$ K ($n_1 = 5$) and $\Theta_{E2} = 700$ K ($n_2 = 8$),[5] which is shown by a solid line in **Fig. 5**. With three $Mn^{2+}$ ($d^5$, S = 5/2) magnetic ions per formula unit, the $C_m(T)$ of $\gamma$-$Mn_3(PO_4)_2$ saturates at the value ~45 J/mol K, which is very close to the estimated value, $S_m = nRln(2S+1)$, where n = 3 (i.e., the number of magnetic ions per formula unit) and S = 5/2 for the $Mn^{2+}$ ions of $\gamma$-$Mn_3(PO_4)_2$ (see the inset of **Fig. 5**).

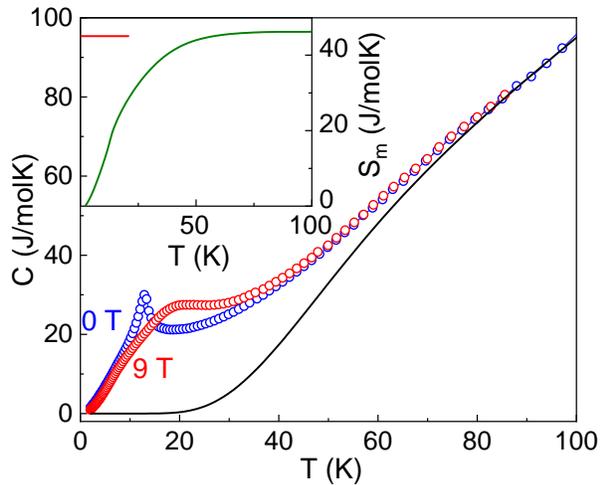

**Fig. 5**. Specific heat of γ-Mn$_3$(PO$_4$)$_2$ taken at $\mu_0H = 0$ T (blue circles) and 9 T (red circles). The black curve represents the phonon contribution. The inset shows the temperature dependence of the magnetic entropy at $\mu_0H = 0$ T, with the red dash indicating the saturation limit of $S_m$.

To check whether the magnetic excitation spectrum preceding the plateau differs from the one at the plateau, we compare the magnetic specific heat $C_m(T) = C(T) - C_{ph}(T)$ taken at $\mu_0H = 0$ T and 9 T in the temperature region of 2 – 6 K, which is far below $T_N = 12.9$ K (**Fig. 6a**). γ-Mn$_3$(PO$_4$)$_2$ consists of ferrimagnetic layers that are weakly coupled antiferromagnetically. If the magnetic excitation spectrum has an energy gap, we can fit the $C_m(T)$ vs. $T$ data using the relationship,[3]

$$C_m(T) = \alpha T^{3/2} \exp(-\Delta/k_B T), \qquad (2)$$

where α is a weighting coefficient and $\Delta$ is the gap in the magnetic excitation spectrum. Using Eq. 2 with $\alpha = 0.68$ J/mol K$^{5/2}$, we obtain $\Delta = 0.5$ K at $\mu_0H = 0$, and $\Delta = 1.4$ K at $\mu_0H = 9$ T. Namely, the gap in the field region of the 1/3-magnetization plateau is quite different from, about 3 times greater than, the one in the low field region far below the plateau. With respect to the $C_m(T) = \alpha T^{3/2}$ relationship (**Fig. 6a**), the $C_m(T)$ curve is lowered more at $\mu_0H = 9$ T than at $\mu_0H = 0$, revealing that the internal energy of the spin lattice is greater at $\mu_0H = 9$ T than at $\mu_0H = 0$.

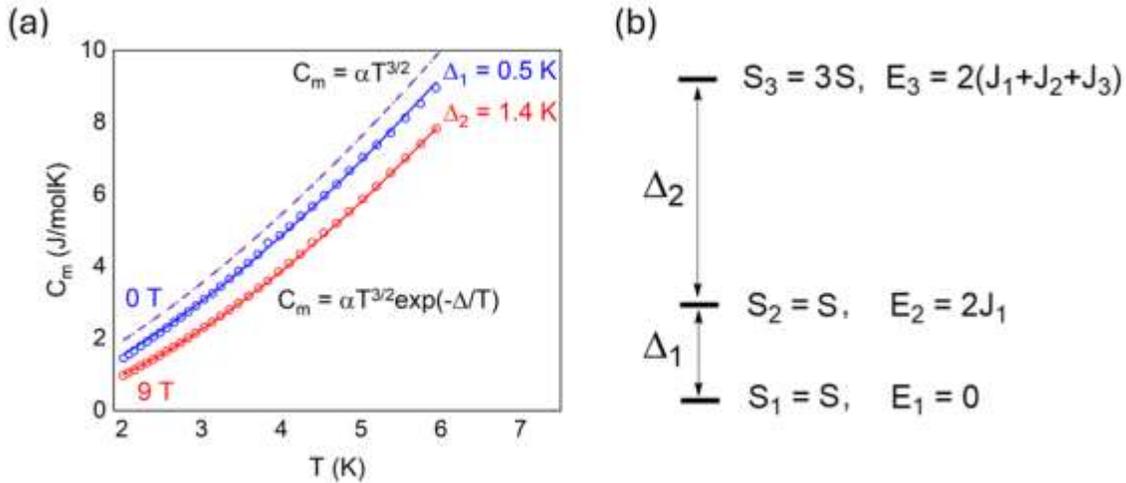

**Fig. 6.** (a) Magnetic specific heat of γ-Mn$_3$(PO$_4$)$_2$ taken at $\mu_0H = 0$ T (blue circles) and 9 T (red circles). The blue and red solid curves represent the fitting relationship $C_m(T) = \alpha T^{3/2} \exp(-\Delta/k_B T)$

with $\alpha$ = 0.68 J/mol K$^{5/2}$. The dash-dotted line represents the relationship $C_m(T) = \alpha T^{3/2}$ with $\alpha$ = 0.68 J/mol K$^{5/2}$. (b) Spins and local energies associated with three magnetic states of a linear trimer Mn$^{2+}$-Mn$^{2+}$-Mn$^{2+}$ embedded in the spin lattice of $\gamma$-Mn$_3$(PO$_4$)$_2$, where $S$ = 5/2, from the viewpoint of spin exchange interactions (see **Fig. 4c – 4f**.)

Under magnetic field $\mu_0 H$, a magnetic species with spin $S$ has the Zeeman energy $E_Z = \mu_0 \mu_B g \vec{S} \cdot \vec{H}$. At zero field, $E_Z$ = 0. Under nonzero field, Zeeman energy induces spin-lattice interactions, which raises the internal energy of the lattice (see below for further discussion from the viewpoint of Le Chatelier's principle). This in turn reduces its specific heat. To a first approximation, such an effect of nonzero Zeeman energy is absent under zero field. This explains why the $C_m(T)$ is lower at $\mu_0 H$ = 9 T than at $\mu_0 H$ = 0. We probe the microscopic cause for this phenomenological observation and its implications by probing the local magnetic excitations available, at a given external magnetic field, for a linear ferrimagnetic trimer embedded in the 3D spin lattice. (Local magnetic excitations were also found to be important in understanding why the ordered spin moments of the Fe$^{3+}$ (S = 5/2) ions in LiFeV$_2$O$_7$, determined by neutron diffraction at very low temperatures, are strongly reduced from the expected value of 5 $\mu_B$.[6]) In the nonzero slope region (below ~6 T) of the magnetization curve, the interlayer magnetic bonds J$_1$ are gradually broken with increasing field until all interlayer magnetic bonds are broken. During this process, the ferrimagnetic trimer units belonging to the N – M regions remain bonded to each other by the inter-trimer bonds J$_2$ at fields around $\mu_0 H$ = 0 T (**Fig. 4e**). Thus, as far as an individual trimer is concerned, the energy gap $\Delta_1$ between after and before the inter-layer magnetic bond breaking is given by $\Delta_1$ = E$_2$ – E$_1$ = 2J$_1$ per trimer (**Fig. 6b**) from the viewpoint of the spin exchanges involved. In the field region of the 1/3-magnetization plateau (i.e., between ~6 and ~23 T), each ferrimagnetic trimer is bonded to four adjacent ferrimagnetic trimers by four inter-trimer bonds J$_2$ (**Fig. 4e**). Beyond ~23 T, each individual ferrimagnetic trimer starts to become fully magnetized as depicted in **Fig. 4f**, which requires the breaking of two inter-trimer bonds J$_2$ and two intra-trimer bonds J$_3$. Thus, the local magnetic excitation energy gap $\Delta_2$ is given by $\Delta_2$ = E$_3$ – E$_2$ = 2(J$_2$ + J$_3$) per ferrimagnetic trimer (**Fig. 6b**) from the viewpoint of the spin exchanges involved. In terms of the calculated spin exchanges (J$_1$ = 1.7 K, J$_2$ = 4.7 K and J$_3$ = 10.5 K [2]), the $\Delta_2/\Delta_1$ ratio is estimated to be ~4.5. Given the rough approximations involved in estimating the two successive local excitations of a trimer, this is in reasonable agreement with the observed ratio of ~2.8 (= 1.4/0.5) determined from the field-dependence of the gaps determined experimentally.

### 3.2.B. Field-dependence of specific heat

Let us now consider the magnetization $M(H)$ and specific heat $C(H)$ of $\gamma$-Mn$_3$(PO$_4$)$_2$ measured as a function of a magnetic field $\mu_0 H$ at 2 K by sweeping $\mu_0 H$ from 0 to 9 T (up-field sweep) as well as from 9 T to 0 (down-field sweep). The magnetization curves $M(H)$ (**Fig. 7a**) evidence a spin-flop transition starting at $\mu_0 H_{sf}$ = 4 T and a 1/3 plateau transition starting at $\mu_0 H_{pl}$ = 5.85 T, which are more clearly recognizable from the d$M$/d$H$ curve (**Fig. 7b**) obtained during the down-field sweep. The transition at 4 T, which is due to the spin-flop transition, is of the first order because the d$M$/d$H$ curve around 4 T is symmetric in shape and hysteretic. The transition at 5.85 T, which is due to the formation of the 1/3-magnetization plateau, is of the second order because the d$M$/d$H$ curve around 5.85 T is non-symmetric and non-hysteretic. Since this transition at 5.85 T is second order in nature, it rules out the transition from an inhomogeneous two-phase state to a homogeneous single-phase state, and vice versa. The 1/3-plateau phase is homogeneous, because it is made up of only the partitioned-out ferrimagnetic triangles. Therefore, the process of field-induced spin-polarization cannot be Model B. To determine whether Model A, C or D is correct for the field-induced spin-polarization, it is necessary to examine if their predictions on magnetic entropy is consistent with the field-dependent specific heat $C(H)$ presented in **Fig. 7c**, which shows that the $C(H)$ vs. $\mu_0 H$ plot is nearly flat in the 0 to ~6 T region (i.e., a nonzero slope region of $M(H)$) but $C(H)$ decreases sharply with field in the region beyond ~6 T (i.e., a magnetization plateau region of $M(H)$). The $C(H)$ obtained for the up-field sweep is not shown, because it is very similar to **Fig. 7c** (see **Fig. S4** in Section S2).

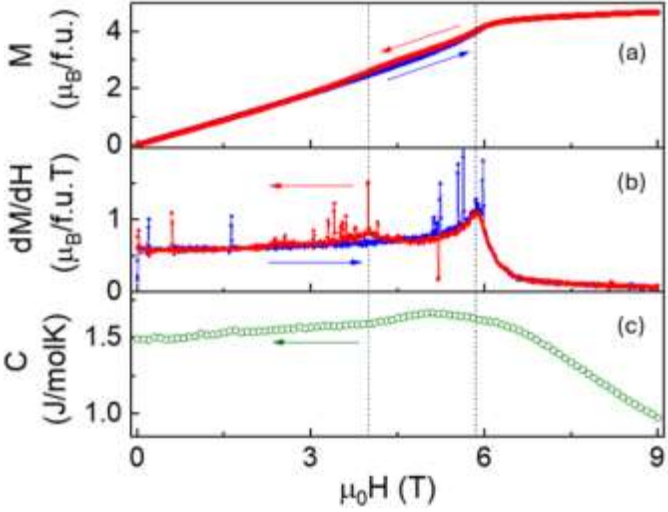

**Fig. 7**. (a) Magnetization $M(H)$, (b) its derivative $dM(H)/dH$ and (c) specific heat $C(H)$ of γ-$Mn_3(PO_4)_2$ measured as a function of $\mu_0H$ at 2 K.

In discussing the temperature-dependent specific heat $C(T) = C_m(T) + C_{ph}(T)$ of a magnet at low temperatures (e.g., 2 K), it is commonly believed that the phonon contribution $C_{ph}(T)$ is "quenched", so $C(T)$ is dominated by the magnon contribution $C_m(T)$. As discussed in the previous section, the $C_m(T)$ depends on the local magnetic energy spectrum of a magnetic species constituting the spin lattice, but not on the magnetic entropy. The field-dependent specific heat, $C(H)$, is the sum of contributions from magnetic entropy, $C_m(H)$, and internal energy, $C_{ph}(H)$, namely, $C(H) = C_m(H) + C_{ph}(H)$. To discuss the field-dependent specific heat $C(H)$ of γ-$Mn_3(PO_4)_2$ (**Fig. 7c**), we recall that the internal energy of a solid involves primarily the vibrations of its constituent atoms around their equilibrium positions, and that the specific heat of a solid increases (decreases) with decreasing (increasing) its internal energy, while it increases (decreases) with increasing (decreasing) its magnetic entropy. In what follows, we first examine the field region of the 1/3-magnetization plateau and then the region preceding the 1/3-plateau.

**Field region of the 1/3-magnetization plateau.** As already discussed, the magnetic entropy in this region remains constant so there is no magnetic entropy contribution to specific heat, i.e., $C_m(H) = 0$. The specific heat $C(H)$, and hence $C_{ph}(H)$, of γ-$Mn_3(PO_4)_2$ decreases steadily as $\mu_0H$ increases (**Fig. 7c**), which implies that the internal energy of γ-$Mn_3(PO_4)_2$ increases steadily with increasing field hence requiring less heat to raise the temperature by a unit degree, which in turn leads to a steady lowering of $C(H)$ with increasing field. At a low temperature (i.e., 2 K in the present case), the lattice vibrations are essentially quenched. This prompts us to probe how a solid at very low temperatures raises its internal energy by increasing the external magnetic field. On a qualitative level, this question can be answered by considering a series of events governed by Le Chatelier's principle. The spin lattice of $\gamma$-Mn$_3$(PO$_4$)$_2$ can counteract the external field in two ways; (1) The weak interlayer magnetic bonds (i.e., the spin exchange J$_1$) are broken to generate the partitioned-out ferrimagnetic layers. The ferrimagnetic units (namely, ferrimagnetic trimers) of a partitioned-out ferrimagnetic layer generate nonzero Zeeman energy $E_Z(H)$, which increases with increasing $\mu_0 H$. (2) The ferrimagnetic units will counteract the increase in $E_Z$ by enhancing the spin-lattice interaction. The latter in turn induces an increase in the vibrational energy of each lattice site by mixing the vibrational excited state into the vibrational ground state of each lattice site, which involves the Boltzmann factor, exp(-$\Delta/k_B T$), where $\Delta$ refers to the energy gap between the two vibrational states in this case. This raises the internal energy of the spin lattice thereby decreasing $C_{ph}(H)$ with field so that the $C(H)$ should decrease with increasing $\mu_0 H$, in agreement with the observation.

**Field region preceding the 1/3-magnetization plateau**. In this field region of the magnetization curve, the $C(H)$ increases very slowly reaching a broad maximum (around 5 T) as $\mu_0 H$ increases from 0 to ~6 T (**Fig. 7c**). This behavior should be a consequence of two competing factors. In terms of the internal energy change induced by the field, $C_{ph}(H)$ should decrease with increasing the field as discussed above. Thus, there must be a factor which increases $C(H)$ with increasing $\mu_0 H$. This factor is most likely magnetic entropy $C_m(H)$ so that $C(H) = C_{ph}(H) + C_m(H)$. Since $C_m(H)$ must increase with field in the 0 to ~6 T region, Model A is ruled out as the process of field-induced spin-polarization, because it predicts that magnetic entropy remains constant. Model B is also ruled out, because it predicts that magnetic entropy decreases with field. This leaves only Model C or D to consider.

In the 0 to ~6 T region, the interlayer magnetic bonds J$_1$ are broken to generate partitioned-out ferrimagnetic layers. Suppose that $\gamma$-Mn$_3$(PO$_4$)$_2$ consists of $L$ ferrimagnetic layers. These layers are antiferromagnetically coupled, so the magnetization increases as the number of partitioned-out ferrimagnetic layers increases. As already discussed, one might approximate the number of different ways, $\Omega(p)$, to choose $p$ partitioned-out ferrimagnetic layers from the $L$ layers by the binomial coefficient $_LC_p$, namely, $\Omega(p) = {}_LC_p$. Then, the associated magnetic entropy $S(p)$ is given by $k_B ln\Omega(p)$. In this approximation, $\Omega(p)$ is a symmetric function with maximum at $p = L/2$, so is $S(p)$. Then, the magnetic entropy would increase when $p$ increases from 0 to $L/2$ (i.e., in the field region from 0 to ~3 T), but it would decrease when $p$ increases from $L/2$ to $L$ (i.e., in the field region from ~3 to ~6 T) (as depicted by Model C in **Fig. 1e**). This does not explain the near flatness of the $C(H)$ vs. $H$ curve in the 0 to ~6 T region with a broad maximum around ~5 T, which indicates that the magnetic entropy contribution is slightly stronger than the internal energy contribution. To explain these features of the $C(H)$ vs. $H$ curve, the magnetic entropy must increase steadily with increasing the field from 0 to ~6 T, as depicted by Model D in **Fig. 1e**, because the internal energy would increase steadily as the field increases from 0 to ~6 T. The correct behavior of the magnetic entropy is not described by $\Omega(p)$, because it neglects the role of the interlayer magnetic bonds in partitioning out ferrimagnetic layers.

For each ferrimagnetic layer to be partitioned out in $\gamma$-Mn$_3$(PO$_4$)$_2$, its interlayer bonds with the two adjacent layers should be broken (**Fig. 4b**). Let us represent the spin lattice with no broken interlayer bonds as depicted in **Fig. 8a**, where each layer is indicated by a black rectangular box, and the unbroken interlayer bonds by solid black lines. We might choose one partitioned-out

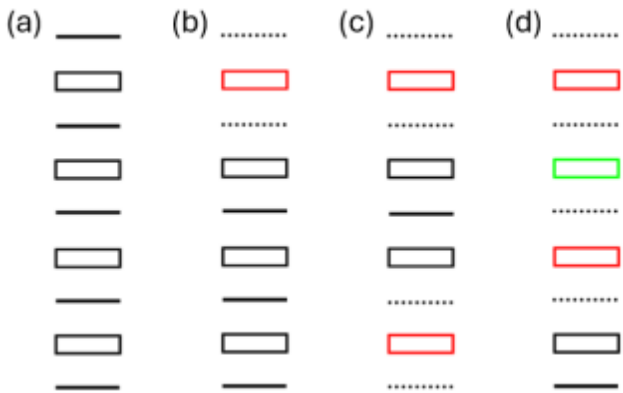

**Fig. 8**. (a) Ferrimagnetic layers antiferromagnetically coupled by interlayer bonds. (b) Formation of one partitioned-out ferrimagnetic layer. (c) Formation of two isolated ferrimagnetic layers. (d) Formation of two isolated ferrimagnetic layers causing the formation of another isolated

ferrimagnetic layer. Each rectangular box represents a ferrimagnetic layer, and the solid and dotted between the layers the unbroken and broken interlayer magnetic bonds, respectively.

ferrimagnetic layer, which is a member of $\Omega(1)$, by representing it with a red rectangular box and its broken interlayer bonds by black dotted lines (**Fig. 8b**). We choose an example of two partitioned-out ferrimagnetic layers, which is a member of $\Omega(2)$, such that the two ferrimagnetic layers are beyond the second nearest neighbors (**Fig. 8c**). If the two chosen layers are second nearest neighbors (**Fig. 8d**), another ferrimagnetic layer becomes partitioned out as indicated by a green rectangular box because its two interlayer bonds are already broken. In other words, one member belonging to $\Omega(2)$ generates a member of $\Omega(3)$. Let us use the notation, $\langle\Omega(p)\rangle$, to refer to the correct number of different ways of producing $p$ partitioned-out ferrimagnetic layers. Then, the associated magnetic entropy is written as $\langle S(p)\rangle = k_B ln\langle\Omega(p)\rangle$. The values of $\langle\Omega(p)\rangle$ can be determined by using the cyclic boundary condition. That is, the $L$ layers are arranged sequentially as $1, 2, 3, \cdots, L-2, L-1, L$ with layer $L$ connected back to layer 1. Note that the interlayer bonds are present between every two adjacent layers. This condition leads to the result $\langle\Omega(L\text{-}1)\rangle = 0$ because all interlayer bonds are broken for any choice of $L-1$ partitioned-out ferrimagnetic layers, so all cases of $\Omega(L\text{-}1)$ end up generating $L$ partitioned-out ferrimagnetic layers. The $\langle\Omega(L\text{-}1)\rangle$ case is excluded from the consideration of magnetic entropy $\langle S(p)\rangle$, because it does not contribute to the total number of possible choices leading to the configurational entropy. It should be pointed out that

$$\sum_{p=0}^{L} \Omega\,(p) = \sum_{p=0}^{L} \langle\Omega(p)\rangle$$

As a simple representative example, we compare the $\Omega(p)$-vs-$p$ and the $\langle\Omega(p)\rangle$-vs-$p$ plots for $L = 6$ (**Fig. 9**). The $\langle\Omega(p)\rangle$-vs-$p$ plot is highly asymmetrical; In general, $\langle\Omega(p)\rangle$ increases with increasing $p$ toward $L$. Thus, the near flatness of the $C(H)$ vs. $H$ curve reflects that $\langle S(p)\rangle$ increases as $p$ increases from 1 to $L$, i.e., with increasing field, while the internal energy increases with field.

In general, the magnetic entropy of any antiferromagnet with or without spin frustration should be described by the modified binomial coefficients $\langle\Omega(p)\rangle$ because the partitioning-out ferrimagnetic fragments generates three different bonding-environments of ferrimagnetic fragments, which are distinguished by the nature of their inter-fragment bonds: (a) Type-A: Partitioned-out fragments possessing only broken inter-fragment bonds. (b) Type-B: Unpartitioned fragments possessing only unbroken inter-fragment bonds. (c) Type-C: Unpartitioned fragments possessing both broken and unbroken inter-fragment bonds, which form the boundary between the partitioned-out and unpartitioned fragments. It is the Type-C fragments that cause the magnetic entropy to depend on the modified binomial coefficients $\langle \Omega(p) \rangle$. This was discussed by considering an AFM chain of antiferromagnetically coupled ferrimagnetic trimers as an example for an antiferromagnet with no spin frustration (see Section S1). Note that, if each ferrimagnetic layer of $\gamma$-$Mn_3(PO_4)_2$ is treated as a ferrimagnetic unit (e.g., a ferrimagnetic linear trimer), then the 3D AFM spin lattice of $\gamma$-$Mn_3(PO_4)_2$ becomes topologically equivalent to an AFM chain of antiferromagnetically-coupled ferrimagnetic units (e.g., an AFM chain of antiferromagnetically-coupled ferrimagnetic linear trimers, **Fig. 1a**).

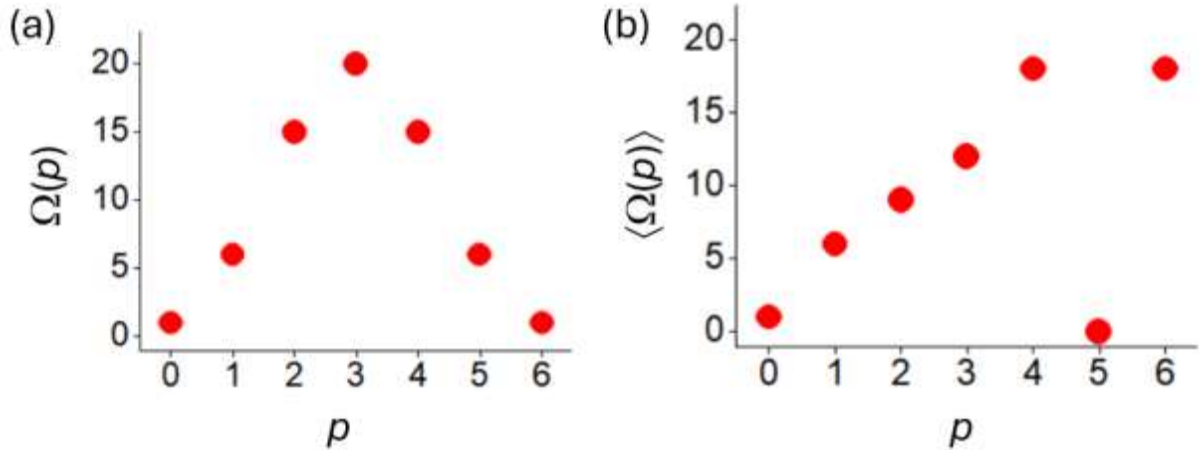

**Fig. 9**. Plots of (a) $\Omega(p)$ vs. $p$ and (b) $\langle \Omega(p) \rangle$ vs. $p$ for the case of $L = 6$.

### 3.2.C. Estimation of the field-dependence of magnetic specific heat

It is important to find experimental support for the contribution of configurational magnetic entropy to the specific heat discussed above. For this purpose, we evaluate the field-dependence of the magnetic specific heat $C_m(H)$ of γ-Mn$_3$(PO$_4$)$_2$ from that of its specific heat $C_{exp}(H)$ measured experimentally by analyzing the field-dependence of the internal energy. In essence, applying magnetic field to a magnet is equivalent to raising its temperature. For an ion of spin $S$ under magnetic field $\mu_0H$, its Zeeman energy $E_Z(H)$ is given by $\mu_B(2S)(\mu_0H)$. Therefore, for an $S = 5/2$ ion under field $\mu_0H$ (in units of Tesla), $E_Z(H) = 3.35\ \mu_0H$. Since the specific heat $C_{exp}(H)$ has two contributions, i.e., $C_{exp}(H) = C_m(H) + C_{ph}(H)$, the field-dependence of $C_m(H)$ can be deduced from the expression, $C_m(H) = C_{exp}(H) - C_{ph}(H)$.

As already mentioned, the field dependence of $C_{ph}(H)$ is determined by that of the internal energy. At any field $\mu_0H$, one might assume that $E_Z(H)$ of a magnetic ion is completely absorbed into the lattice as thermal energy. Then, $E_Z(H)$ amounts to the field-induced energy added to the internal energy, namely, it is equivalent to heating by the amount of $E_Z(H)$. The $E_Z(H)$ vs. $\mu_0H$ relationship for γ-Mn$_3$(PO$_4$)$_2$ can be deduced from the magnetization curve in the field region of the 1/3-magnetization plateau, where there is no contribution of magnetic entropy to the specific heat so that the almost linear decrease of $C_{exp}(H)$ with increasing $\mu_0H$ means an almost linear increase in the internal energy. Since $E_Z(H)$ increases linearly with $\mu_0H$, it is reasonable to suppose that $C_{ph}(H)$ decreases linearly with $\mu_0H$. Thus, the near linear part of the $C_{ph}(H)$ vs. $\mu_0H$ curve in the 5.8 to 9.0 T region of **Fig. 7c** can be fitted by a linear equation, $C_{ph}(H) = a - bE_Z(H) = a - 3.35b(\mu_0H)$ with $a$ and $3.35b$ as fitting parameters. Then, we obtain $C_{ph}(H) = 3.19 - 0.25(\mu_0H)$, which shows that the $C_{ph}(H)$ decreases at the rate of 0.25 J/(molK) per Tesla. Thus, the $C_{ph}(H)$ in the 0 to ~6 T region is given by $-0.25(\mu_0H)$ because this effect vanishes at $\mu_0H = 0$. Consequently, the $C_m(H)$ in the 0 to ~6 T region is written as $C_m(H) = C_{exp}(H) - C_{ph}(H) = C_{exp}(H) + 0.25(\mu_0H)$.

The $C_m(H)$ vs. $\mu_0H$ resulting from this equation $C_m(H) = C_{exp}(H) + 0.25(\mu_0H)$ is presented in **Fig. 10**. This plot shows that the magnetic specific heat increases steadily with increasing field from 0 to ~6 T, which means that the magnetic entropy increases steadily with increasing field from 0 to ~6 T. This result is in support of our conclusion that the magnetic entropy of γ-Mn$_3$(PO$_4$)$_2$ in the 0 to ~6 T region is described by $\langle S(p) \rangle = k_B ln \langle \Omega(p) \rangle$.

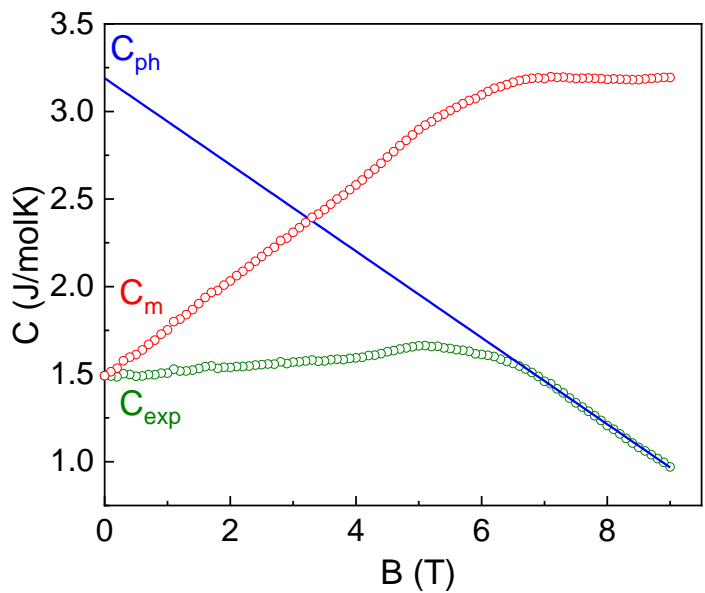

**Fig. 10**. The magnetic specific heat $C_m(H)$ of γ-Mn$_3$(PO$_4$)$_2$ as a function of $\mu_0 H$ in the 0 to 9 T region. For comparison, the $C_{exp}(H)$ vs. $\mu_0 H$ plot is also shown.

## 4. Discussion

The increase in magnetic entropy of γ-Mn$_3$(PO$_4$)$_2$ with increasing field in the 0 to ~6 T region is explained by the increase in configurational magnetic entropy, which requires the heterogeneity of magnetic entropy distribution. However, the field-induced phase transition at 5.85 T is second order, requiring that the states below 5.85 T be homogeneous. The latter is achieved when the two magnetic phases of different magnetic entropy shift their spin sites rapidly and dynamically.

Field-induced partitioning of an AFM spin lattice into ferrimagnetic fragments involves field-induced breaking of certain magnetic bonds leading to a heterogeneous change in magnetic entropy and hence generating two magnetic phases of different magnetic entropy, which dynamically shift their spin sites such that all spin sites appear homogeneous for magnetization measurements. Field-induced partitioning of a spin lattice into ferrimagnetic fragments raises their Zeeman energy at the expense of reducing the magnetic entropy of individual magnetic fragments.

Formation of the smallest ferrimagnetic fragments would be energetically most favorable because it maximizes the configurational magnetic entropy and hence the energy stabilization by $-T\Delta S$.

The magnetization plateau phenomenon in spin frustrated antiferromagnets has been studied over the years using spin Hamiltonians both with[7] and without[8,9] including spin-lattice coupling terms. The implicit assumption of all such studies is that total energy changes in antiferromagnets are well described by considering only their enthalpy changes. Our study shows that entropy changes, and hence free energy changes, are crucial in describing total energy changes at low temperatures.

As discussed in Section 2.1B, it is energetically more favorable to surround a ferrimagnetic triangle with spin-unpolarized triangles rather than with ferrimagnetic triangles by $(1/25)J$ per i-j contact between ferrimagnetic triangles. The two, three and four adjacent ferrimagnetic triangles in **Fig. 2d**, **2f** and **2h** make three, seven and 11 such i-j contacts, respectively, leading to the destabilization of $(3/50)J$, $(7/75)J$ and $(11/100)J$ per ferrimagnetic triangle, respectively. The ferrimagnetic triangles created by the field prefer to be surrounded by spin-unpolarized triangles than by ferrimagnetic triangles. Thus, in the early stage of magnetization where there are many more spin-unpolarized triangles than partitioned-out ferrimagnetic triangles, the ferrimagnetic triangles would stay separated instead of aggregating, and a partitioned-out ferrimagnetic triangle can be anywhere within a region of spin-unpolarized triangles with equal stability. It would be interesting to detect the presence of such isolated ferrimagnetic triangles in a layered antiferromagnet with a trigonal spin lattice by surface magneto-optic Kerr effect (SMOKE) measurements[10] with magnetic field applied perpendicular to the layer or by Mössbauer spectroscopy measurements.[11] An ideal trigonal antiferromagnet for such measurements would be $RbFe(MoO_4)_2$.[1,12]

The field-induced formation of partitioned-out ferrimagnetic fragments in a triangular spin lattice is analogous to that of vortices in a type-II superconductor under magnetic field,[13] although the latter is a macroscopic phenomenon in type-II superconductors whereas the former is a microscopic phenomenon in magnetic insulators. The ferrimagnetic triangles correspond to the quantized flux tubes while the spin-unpolarized triangles surrounding each ferrimagnetic triangle to a vortex of superconducting current surrounding each quantized flux tube. The partitioned-out ferrimagnetic triangles tend to remain separated under magnetic field. Similarly, the vortices of a type-II superconductor under magnetic field remain separated.

In our discussion of the $C_m(T)$ vs. $T$ relationship, the energy gap $\Delta$ of the magnetic excitation spectrum provides information about the local magnetic states of the magnetic species constituting the spin lattice. From the viewpoint of the Boltzmann statistics, the local excitation amounts to mixing the excited state (i.e., unoccupied) state into the ground state (i.e., occupied) state, which is separated by the energy gap $\Delta$, by the amount of the Boltzmann factor exp(-$\Delta/k_B T$). This explains the occurrence of the exponential decay term in Eq. 2. This picture of local magnetic excitations arising from the temperature dependence of $C_m(T)$ provides support for the occurrence of two phases and the heterogeneous change in magnetic entropy in the two-phase regions of a magnetization curve. As already pointed out, ferrimagnetic species under field can increase the internal energy by a series of events governed by Le Chatelier's principle, namely, increase in the Zeeman energy to enhance the spin-lattice interactions, which in turn increases the vibrational energy thereby raising the internal energy. The last step would involve the mixing of the vibrational excited states into the vibrational ground states via the Boltzmann factor.

## 5. Concluding remarks

Magnetization plateaus of various antiferromagnets are readily explained by the supposition[1] that an antiferromagnet counteracts the field by partitioning its spin lattice into ferrimagnetic fragments and hence absorbing Zeeman energy according to Le Chatelier's principle. To find a theoretical basis for the supposition, we analyzed how external magnetic fields influence the magnetic entropies of antiferromagnets exhibiting the magnetization plateau phenomenon by examining a trigonal spin lattice with nearest-neighbor AFM spin exchange as an example with spin frustration as well as an AFM chain of antiferromagnetically coupled ferrimagnetic linear trimers as an example with no spin frustration, and verified our conclusions by measuring the temperature- and field-dependent specific heats of $\gamma$-$Mn_3(PO_4)_2$. Our main conclusions are summarized as follows:

(1) The nature of the magnetic structure of an antiferromagnet depends on the region of its magnetization curve. In a nonzero slope region, the magnetic structure is heterogeneous because it consists of partitioned-out and unpartitioned ferrimagnetic fragments. In a magnetization plateau region, the magnetic structure is homogeneous because it consists of identical magnetic fragments.

(2) In a single-phase region, the magnetic entropy is independent of magnetic field. In a two-phase region, the magnetic entropy increases with increasing field as described by the modified binomial coefficients, i.e., $\langle S(p) \rangle = k_B ln \langle \Omega(p) \rangle$, because the spin sites of the partitioned-out and unpartitioned ferrimagnetic fragments shift their spin sites dynamically, and because the field-induced partitioning generates ferrimagnetic fragments possessing both broken and unbroken inter-fragment bonds in addition to those possessing only broken inter-fragment bonds and those with only unbroken inter-fragment bonds.

(3) The field-induced breaking and hence the field-induced partitioning of an antiferromagnetic spin lattice into ferrimagnetic fragments are the time-averaged results of all allowed spin arrangements that occur repeatedly in the static magnetization measurements. For a broken magnetic bond, the effective spin exchange is zero as if the bond is physically broken.

(4) The specific heat $C(H)$ between 0 and 9 T measured at 2 K is nearly flat between 0 and ~6 T (i.e., a two-phase region) with a broad maximum around 5 T but decreases sharply as $\mu_0 H$ increases beyond ~6 T (i.e., a single-phase region). These observations reflect that the internal energy contribution to $C(H)$ decreases with increasing field in both regions while the magnetic entropy contribution to $C(H)$, which occurs only in the two-phase region, increases with increasing field from 0 to ~6 T.

(5) The temperature-dependent magnetic specific heat $C_m(T)$ measured for $\gamma$-Mn$_3$(PO$_4$)$_2$ between 2 – 6 K leads to the energy gap $\Delta = 0.5$ K when measured under $\mu_0 H = 0$, but to the energy gap $\Delta = 1.4$ K when measured under 9 T. This field-dependence of is related to the Boltzmann factors associated with the local magnetic excitations of individual ferrimagnetic trimers embedded in the AFM spin lattice of $\gamma$-Mn$_3$(PO$_4$)$_2$.

(6) Le Chatelier's principle offers a qualitative explanation for the sequence of events caused by an external magnetic field in antiferromagnets that show magnetization plateaus. These events include an increase in the field leading to the partitioning of a spin lattice into ferrimagnetic fragments, which then results in increased spin-lattice interactions, increased vibrational energy, increased internal energy, and a subsequent decrease in specific heat.

**Acknowledgements**

The authors would like to thank Pavel Maksimov, Larisa V. Shvanskaya, Reinhard K. Kremer and Alexander N. Vassiliev for valuable discussions during the completion stage of this work. The research at KHU was supported by Basic Science Research Program through the National Research Foundation of Korea (NRF) funded by the Ministry of Education (RS-2020-NR049601). OSV acknowledges support of the research carried out within the framework of the scientific program of the National Centre of Physics and Mathematics under the project "Research in strong and superstrong magnetic fields" and support from the Ministry of Science and Higher Education of Russia in the framework of the Program of Strategic Academic Leadership "Priority 2030" (MISIS Strategic Project Quantum Internet).

**Conflicts of interest**

There are no conflicts to declare.

**Data availability**

The data supporting this article has been included as part of the Supplementary Information.

**Supplementary material**

Supplementary Section S1 for magnetic bond breaking in antiferromagnets with no spin frustration with Fig. S1 and Fig. S2 (pdf). Section S2 for additional supplementary figures Fig. S3 and Fig. S4 (pdf).

TOC Graphic

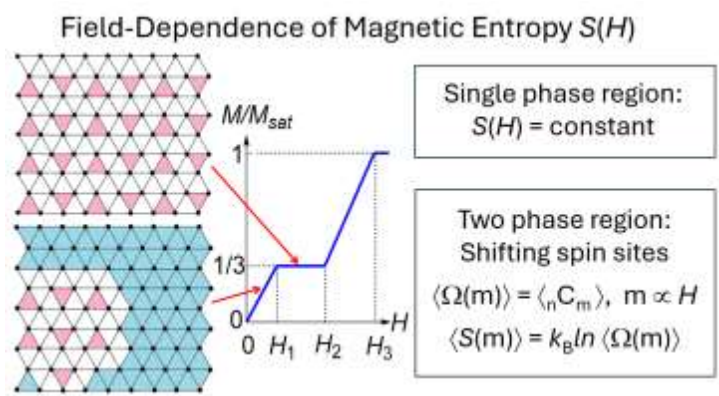

Supporting Information

for

**Le Chatelier's Principle and Field-Induced Change in Magnetic Entropy Leading to of Spin Lattice Partitioning and Magnetization Plateau**

Myung-Hwan Whangbo[1,*], Hyun-Joo Koo[2] and Olga S. Volkova[3,*]

[1] Department of Chemistry, North Carolina State University, Raleigh, NC 27695-8204, USA

[2] Department of Chemistry, Research Institute for Basic Sciences, Kyung Hee University, Seoul 02447, Republic of Korea

[3] Department of Low Temperature Physics and Superconductivity, Lomonosov Moscow State University, Moscow 119991, Russia

mike_whangbo@ncsu.edu

os.volkova@yahoo.com

## Section S1. Magnetic Entropy and Field-induced partitioning of an antiferromagnet with no spin frustration

A chain of antiferromagnetically coupled ferrimagnetic trimers, described by the intra- and inter-trimer spin exchanges J and J′, respectively, is a simple, representative example of antiferromagnets with no spin frustration. The spin arrangement of this AFM chain can be represented in two equivalent ways as depicted in **Fig. S1**, where the unshaded and shaded spheres refer to the up- and down-spins at the spin sites of this chain, respectively. Consequently, each spin site can have either an up-spin or a down-spin. Thus, when the field partitions out a ferrimagnetic fragment (i.e., a ferrimagnetic trimer), the spin-unpolarized parts of the chain adjacent to it have two possible spin arrangements, as shown in **Fig. 2Sa – Fig. 2Sd**, where the spin site $i$ of the partitioned-out ferrimagnetic trimer bridged to the spin site $j$ of the spin-unpolarized parts of the chain were defined. Then, the possible numbers of up- and down-spin occurrences at the $i$ and $j$ sites are as follows:

$p_i\uparrow = 1$, and $p_i\downarrow = 0$,

$p_j\uparrow = 1$, and $p_j\downarrow = 1$.

Thus, across each magnetic bond $i$–$j$, there occurs one AFM and one FM spin arrangement. Effectively, then, there exists no spin exchange in the magnetic bond $i$–$j$, i.e., this magnetic bond is broken. If two ferrimagnetic trimers were to be adjacent to each other, the spin exchange between them would be FM with strength of J′, so this interaction is less stable than that of a ferrimagnetic trimer with the spin-unpolarized part. Therefore, spin-polarized trimers in the AFM chain prefer to stay separated rather than being agglomerated.

Suppose that the AFM chain is made up of $n$ antiferromagnetically coupled ferrimagnetic trimers. Then, the number of field-induced spin-polarized trimers, $m$, increases with increasing the magnetic field. The configurational magnetic entropy for the case of $m$ spin-polarized trimers is given by $k_B ln \langle \Omega(m) \rangle$, where $\langle \Omega(m) \rangle$ is the modified binomial distribution coefficient $\langle {}_n C_m \rangle$ defined in the text. There are two reasons for this conclusion: (a) To partition out a ferrimagnetic trimer, it is necessary to break two magnetic bonds it makes, one on its left side and the other on its right side. (b) When the process of partitioning out ferrimagnetic fragments, there are three kinds of ferrimagnetic fragments, namely, those possessing only the broken inter-fragment bonds, those possessing only the unbroken inter-fragment bonds, and those possessing a broken and an unbroken inter-fragment bond.

**Fig. S2e** depicts the spin configuration of the chain where all inter-trimer magnetic bonds are broken. This state is achieved when the external field reaches a certain value, $\mu_0 H_1$. When the field increases from $\mu_0 H_1$, the magnetization does not change unless it becomes strong enough (say, $\mu_0 H_2$) to break two intra-trimer bonds (2J) of a ferrimagnetic trimer to turn it into a fully polarized (i.e., FM) trimer. Thus, the magnetization remains constant at $M_{sat}/3$ in the field region between $\mu_0 H_1$ and $\mu_0 H_2$. As the field increases from $\mu_0 H_2$, the magnetization increases gradually, as ferrimagnetic trimers begin to become FM one by one (**Fig. S2f**) until all trimers become FM (**Fig. S2g**) at a certain field, $\mu_0 H_3$. The magnetization curve expected for the AFM chain is illustrated in **Fig. S2h**.

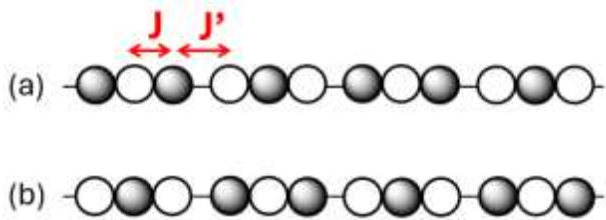

Fig. S1. Two equivalent AFM arrangements for a chain of antiferromagnetically coupled ferrimagnetic trimers, where unshaded and shaded spheres represent the up- and down-spins, respectively.

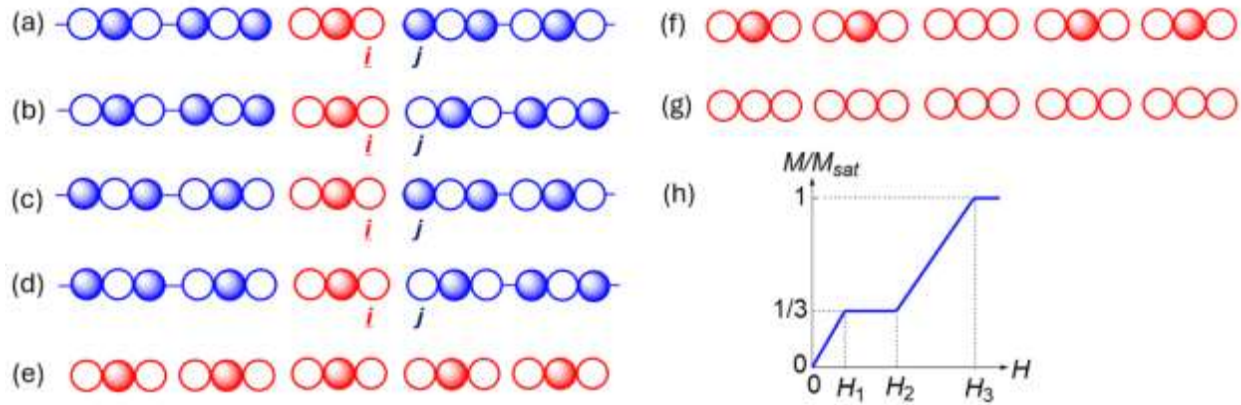

Fig. S2. Field-induced changes in an AFM chain of antiferromagnetically coupled ferrimagnetic trimers, for which the intra- and inter-trimer spin exchanges (J and J′, respectively) are both AFM with J > J′. Here the unshaded and shaded spheres represent the up- and down-spins: (a – d) Four possible spin arrangements of an AFM chain of antiferromagnetically coupled ferrimagnetic trimers containing a partitioned-out ferrimagnetic trimer (represented by red spheres). The spin-unpolarized parts are denoted by blue spheres. The two spin sites associated with each broken magnetic bond are labeled as $i$ and $j$, referring to the spin sites of the ferrimagnetic trimer and the spin-unpolarized part, respectively. (e) Chain of ferrimagnetic trimers in which all inter-trimer bonds are broken. (f) Ferrimagnetic trimers surrounding a fully-spin-polarized (i.e., FM) triangle. (g) A chain made up of FM trimers. (h) The magnetization curve expected for the AFM chain of antiferromagnetically coupled ferrimagnetic trimers.

**Section S2. Additional figures**

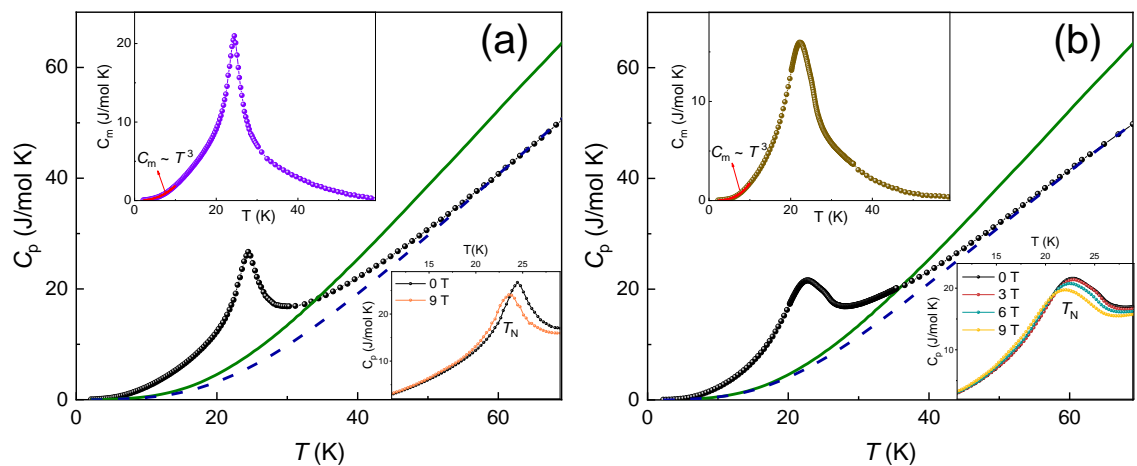

Fig. S3. Temperature dependence of the specific heat measured for (a) $K_2Ni_2TeO_6$ and (b) $Li_2Ni_2TeO_6$.

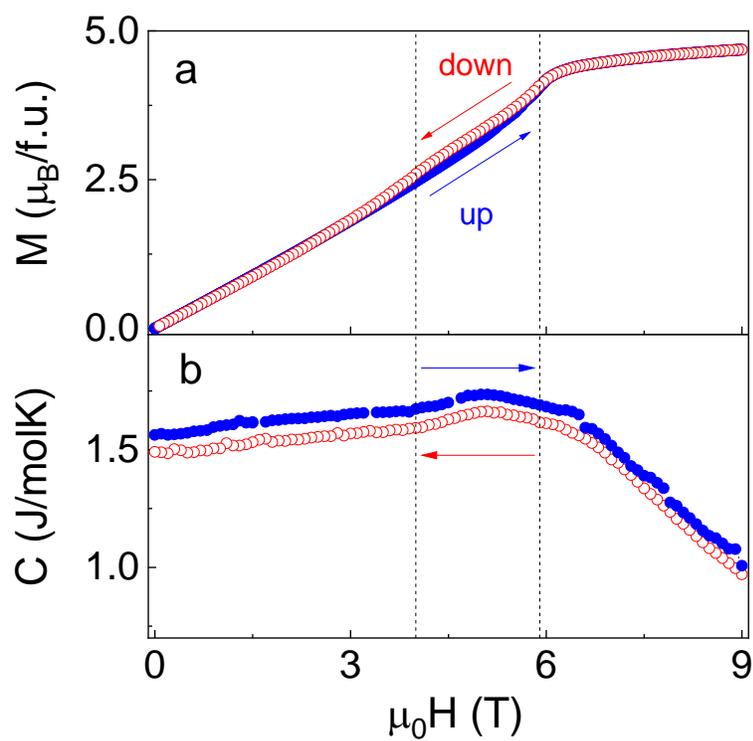

Fig. S4. Field dependences of magnetization (a) and specific heat (b) in γ-Mn₃(PO₄)₂ at ramping magnetic field up and down at 2 K.